\documentstyle[multicol,aps,amssymb,epsfig,array,tabularx]{revtex}
\include{rotating}

\newcommand{\beq}{\begin{equation}}
\newcommand{\eeq}{\end{equation}}
\newcommand{\beqa}{\begin{eqnarray}}
\newcommand{\eeqa}{\end{eqnarray}}
\newcommand{\beqano}{\begin{eqnarray*}}
\newcommand{\eeqano}{\end{eqnarray*}}
\newcommand{\bsubeqa}{\begin{subeqnarray}}
\newcommand{\esubeqa}{\end{subeqnarray}}

\newcommand{\refp}[1]{(\ref{#1})}

\begin{document}
\title{Synchronization of degree correlated physical networks}
\author{M. di Bernardo${}^*$, F. Garofalo${}^*$, F. Sorrentino${}^{*\dag}$}
\address{${}^*$Department of Systems and Computer Science\\
University of Naples Federico II, Naples 80125,
Italy}
\address{${}^\dag$ Corresponding author. E-mail: {\tt
fsorrent@unina.it}}
\date{\today}
\maketitle
\begin{abstract}
We propose that negative degree correlation among nodes in a
network of nonlinear oscillators, often detected in real world
networks, is motivated by its positive effects on
synchronizability. In so doing, we use a novel methodology to
characterise degree correlation based on clustering the network
vertices in $p$ classes according to their degrees. Using this
strategy we construct networks with an assigned power law
distribution but changing degree correlation properties. We find
that the network synchronizability improves when the network
becomes dissassortative, i.e. when nodes with low degree are more
likely to be connected to nodes with higher degree. Our numerical
observations are confirmed by the analytical estimates found in
this letter using an innovative approach based on the use of graph
theoretic results.
\end{abstract}
\begin{multicols}{2}
Networks of oscillators abound in physics, biology and
engineering. Examples include communication networks, sensor
networks, neuronal connectivity networks, biologic networks and
food webs. Under certain conditions such networks are known to
synchronize on a common evolution with all the oscillators
exhibiting the same asymptotic trajectories. % Synchronization is animportant feature,
Moreover synchronization was observed to play an important role in
many different problems of a most diverse nature (physical,
ecological, physiological, meteorological, to name a few); see for
example
\cite{KuraBOOK,WinBOOK,Er:Ko1,Er:Ko2,Wang,exploring,Str1,Str2}.

Recently, it has been proposed that the network topology, i.e. the
way in which the oscillators are mutually coupled between
themselves, has an important effect on its synchronization
properties. In \cite{Ba:Pe02}, it was shown that the eigenratio
$R=\frac{\lambda_N}{\lambda_2}$ between the highest eigenvalue
$\lambda_N$ and the lowest eigenvalue $\lambda_2$ of the Laplacian
associated to the network structure is an essential measure of the
network synchronizability, i.e. the smaller the eigenratio, the
larger the interval of the values of the coupling gain, say
$\sigma$, for which the stability of the synchronous state is
achieved. It is therefore important to characterise how the
network topological features affects the Laplacian eigenratio. For
example, scale free networks, which are common in nature, were
found to show better synchronizability for increasing value of the
power law exponent \cite{Ni:Mo}, \cite{paradox}.

Another important topological property of physical and biological
networks is that often their nodes show preferential attachment to
other nodes in the network according to their degree
\cite{New02Ass}, \cite{New03Mix}. According to this property,
networks are said to exhibit {\em assortative} mixing (or positive
correlation) if nodes of a given degree tend to be attached with
higher likelihood to nodes with similar degree. (Similarly {\em
disassortative} networks are those with nodes of higher degree
more likely to be connected with nodes of lower degree.)

The presence of correlation has been detected experimentally in
many real-world networks. Interestingly, from the analysis of real
networks, it was noticed that social networks are characterized by
positive correlation, while physical and biological networks show
typically a disassortative structure \cite{New02Ass}. For
instance, in \cite{Va:PaCORR}, Internet was found to exhibit
disassortative mixing at the AS level. A pressing open problem is
to understand why negative correlation is an emerging property of
physical and biological network.

In this Letter, we study the effects of degree correlation on the
network synchronizability properties both analytically and
numerically. Our main finding is that disassortative networks of
oscillators synchronize better. These findings have immediate
theoretical and experimental relevance for understanding the
properties of real-world networks. Because of its effects on
synchronizability, we conjecture that disassortative mixing has
played the role of a self organizing principle in leading the
formation of many physical networks as the Internet, the World
Wide Web, protein interactions, neural and metabolic networks.

\begin{figure}[t]
\begin{center}
\epsfig{width=0.50\textwidth, file=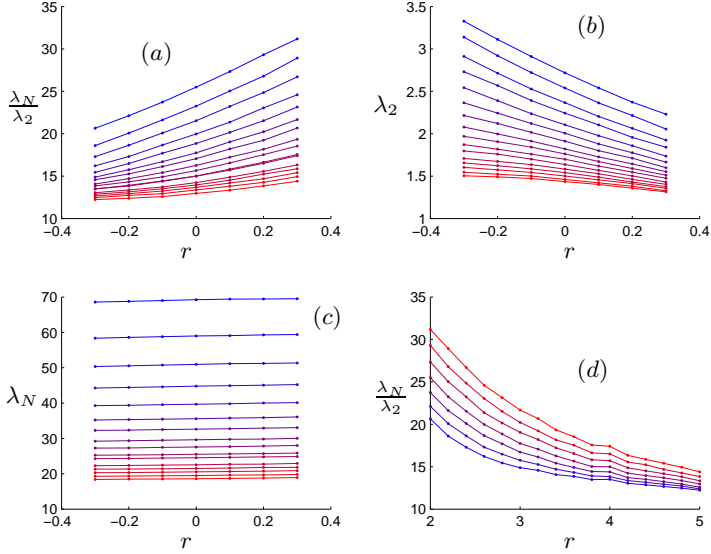}
\begin{picture}(0,0)(0,0)
\put(-70,5){\small ${r}$} \put(-70,115){\small $r$}
\put(-136,60){\small $\lambda_N$} \put(-136,170){\small
$\frac{\lambda_N}{\lambda_2}$} \put(75,5){\small ${r}$}
\put(75,115){\small $r$} \put(3,60){\small
$\frac{\lambda_N}{\lambda_2}$} \put(3,170){\small $\lambda_2$}
\put(-85,190){\small ${(a)}$} \put(-20,90){\small ${(c)}$}
\put(80,200){\small ${(b)}$} \put(80,70){\small ${(d)}$}
\end{picture}
\end{center}
%\centerline{\bf (a) \hspace{5.5cm} (b)}
\caption{\label{fig:R} \narrowtext \small Synchronizability of
degree correlated scale free networks of size $10^3$nodes. %,$k_{min}=5$.
Behavior of the eigenratio $\lambda_{N}/\lambda_{2}$ (a), of the
second lowest eigenvalue $\lambda_2$ (b) and of the highest one
$\lambda_N$ (c), as functions of the correlation coefficient $r$
defined in [11], for $\gamma$ varying from 2 (top line) to 5
(bottom line) in steps of 0.2. (d) The eigenratio as function of
$\gamma$ as varying the correlation coefficient $r$  from $-0.3$
(bottom line) to $0.3$ (top line) in steps of $0.1$. %All the plotsare averaged over $10^2$ realizations.
The lines are guided by the eye.}
\end{figure}

We apply our analysis to scale-free networks in which the
probability of finding a node having degree $k$ scales as
$k^{-\gamma}$. The network generation model is the
\emph{configuration model} as described in \cite{Ne:St01}
\cite{Ni:Mo}. The overall results are summarized in Fig.
\ref{fig:R}(a) where the effects of varying the degree correlation
(measured through the index $r$ defined below) on the Laplacian
eigenratio $R$ are shown for different values of the degree
distribution exponent $\gamma$. For all values of $\gamma$, we
observe a reduction of $R$ for decreasing values of $r$. This
means that disassortative mixing enhances the network
synchronizability. Interestingly, as depicted in Fig.
\ref{fig:R}(b) and Fig. \ref{fig:R}(c),  we observe that, under
variations of the correlation parameter, the changes in $R$ seem
to be mainly due to variations of $\lambda_2$ while $\lambda_N$,
the largest eigenvalue of the Laplacian, is found to be
practically independent from $r$. As discussed in \cite{Ni:Mo}, in
the case of uncorrelated networks ($r=0$), synchronizability
improves for increasing values of $\gamma$.
We find that, as shown in Fig. \ref{fig:R}(d), this is still the
case when degree correlation is introduced. A consistent decrease
of the values of $R$ for the same value of $\gamma$ is observed
when the network is disassortative, i.e. $r<0$.

Historically algebraic graph theory has been widely applied to
describe structure and functions of networks. Here we will give
evidence to the fact that tools from algebraic graph theory can be
conveniently used for the analysis of network dynamical
properties, such as the network {\em synchronizability}.
%when the network dynamical properties, such as the network synchronizability are considered. %when the dynamic over the network is considered
The analytical framework we propose to capture the essence of the
observed phenomenon is based on the following steps: (i) we
introduce a novel index to estimate network assortativity and use
such a quantity to construct networks with fixed degree
distribution but different (desired) degree correlation
properties; (ii) using Cheeger inequalities from algebraic graph
theory, we derive novel bounds on the Laplacian eigenvalues of
interest and hence new analytical bounds on the eigenratio $R$;
(iii) using the new bounds, we validate our numerical
observations.

To start, we assume the network consists of $N$ identical
oscillators coupled through the edges of the network itself
\cite{Ba:Pe02}, \cite{Ni:Mo}, \cite{paradox}. Moreover, we suppose
each oscillator is characterized by its own dynamics, $\{ x_i(t),
i=1...N\}$, described by a nonlinear vector field, say $f=f(x)$,
perturbed by the output function of its neighbors represented by
another nonlinear term, say $h=h(x)$. The equations of motion for
each oscillator can then be given as follows:
\begin{equation}
\frac{dx_i}{dt}=f(x_i)-\sigma \sum_{j=1}^{N} {\mathcal L}_{ij}
h(x_j), \qquad i=1,2,...N, \label{eq:net}
\end{equation}
where $\sigma$ is the overall coupling strength, and ${\mathcal
L}=\{{\mathcal L}_{ij}\}$ is the Laplacian associated to the
network topology \cite{Mohar1}. Given a network of interest, we
start by dividing the network vertices in $p$ classes such that
each one contains $n_1,
n_2...n_p$ nodes and $\sum_i n_i=N$. %We order the vertices in each
%class $1,2...p$ according to the respective degree, that is ineach class ,
In particular, we assign network vertices to each class in order
of increasing (decreasing) degree. Hence, if we label as $k_i$ the
mean degree of vertices belonging to the $i$-th class, then $k_i <
k_{i+1}$ ($i=1,2...,p-1$). %such that the mean degree $k_i$ of the
%vertices belonging to each class satisfies the relation $k_1 < k_2
%< k_p$. %Note that a particular case is the one in which we
%consider exactly $p$ classes as the number of different degrees of
%the network vertices, and we group together in each class all the
%vertices having the same degree.

The probability, $p(i)$, that a generic vertex belongs to class
$i$ is then given by $p(i)=n_i/N$. Thus, considering all the
vertices in each class $i$ as they had mean degree $k_i$, we can
extend the usual definition of degree distribution, assuming that
$p(k_i)=n_i/N$. Analogously the probability of finding a vertex
belonging to class $i$ at the end of a randomly chosen edge within
the network is given by $q(i)=n_i k_i / \sum_i n_i k_i$.

Then, following \cite{New02Ass}, the presence of degree correlation with %(we propose that)
respect to the subdivision in $p$ classes proposed above, can be
estimated by using a new coefficient, $r_p$ defined as:
\begin{equation}
\label{tre} r_p =\frac{{\mathbf{k}}^T ({\mathbf{E}} - {\mathbf{q}}
{\mathbf{q}}^T) \mathbf{k}}{ {\sigma^2_q}},
\end{equation}
where $\sigma_q$ is the standard deviation of the distribution
$q_k$, ${\mathbf{k}}={\pmatrix{k_1,  k_2, ..., k_p}^{T}}$,
${\mathbf{q}}={\pmatrix{q_1,  q_2, ..., q_p}^{T}}$ and
${\mathbf{E}}=\{e_{ij}\} \in {\mathbb{R}}^{{p} \times {p}}$, with
$e_{ij}$ being the probability that a randomly chosen edge in the
network connects nodes belonging to class $i$ and $j$
%As discussed in further details in \cite{ieeepaper}, $r_p$
%provides an estimate of the correlation degree of the network but
%differs from the correlation index $r$ %\refp{eq:r}
%defined in \cite{New02Ass}, \cite{New03Mix}, which bases on the % In particular the index is based on the
%exact degree of the vertices at each of the links
%endpoints in the network. The quantity defined in \refp{tre}
%instead accounts only for whether connections in the networks are
%between nodes belonging to different groups or classes (here we
%suppose each class to be characterized by a certain mean degree,
%but the results we derive can be generalized to whatever a
%subdivision of the network vertices). Nonetheless we found the new
%coefficient to be conceptually equivalent to the one in
%\cite{New02Ass}, \cite{New03Mix} in the sense that it provides a
%reliable indication of the assortative - disassortative nature of
%the network under investigation (
(note that in the particular case where each class contains
exactly only the vertices having a given degree, $r_p$ coincides
with the coefficient $r$ as defined in \cite{New02Ass},
\cite{New03Mix}.)
%the two definitions are practically identical; see \cite{ieeepaper} for further details).% and a comparison between the correlation measure given by
%the two indices $r$ and $r_p$)

From \refp{tre}, it is possible to derive the distribution of
edges among the network vertices as a function of $r_p$ as
follows:
\begin{equation}
\label{EE} {\mathbf{E}}={\mathbf{q}\mathbf{q^T}}+ {r_p} {\sigma^2_q}
\mathbf{M},
\end{equation}
where $\mathbf{M}$ is a symmetric matrix having all row sums equal
to zero and appropriately normalized such that $\mathbf{k^T}
\mathbf{M} \mathbf{k} =1$. Specifically, we can express
${\mathbf{M}}$ as follows:
%as follows:%=\frac{\mathbf{m}\mathbf{m^T}}{z}$, where $\mathbf{m}$ is a vector having
\begin{eqnarray*}
{\mathbf{M}}=\frac{{\mathbf{m}\mathbf{m}}^T}{({\mathbf{k}}^T {\mathbf{m}})^2},  %\qquad
%\mathbf{m}=\pmatrix{ m_1 \cr m_2 \cr ...\cr m_p},
\end{eqnarray*}
where ${\mathbf{m}}={\pmatrix{m_1,  m_2, ..., m_p}^{T}}$ is a
vector such that $\sum_i m_i=0$ (for instance we can choose $m$
such that
$m_i \leq m_{i+1}$ for $i=1,2...,p-1$ %(such a choice is made
in order to have a convenient form of the matrix $\mathbf{M}$ with positive values near the main diagonal
and negative values far away from it).%; $z$ is a normalization constant equal to $({\mathbf{k m}}^T)^2$.

As explained in \cite{ieeepaper}, it is possible to devise a
strategy similar to the one presented in \cite{New02Ass},
\cite{New03Mix}, using the new coefficient $r_p$, to generate
networks with a given degree distribution and a desired value of
the degree correlation coefficient $r_p$. Such a strategy was used
to carry out the computations depicted in Fig. \ref{fig:R}.

\begin{figure}[tbh]
\begin{center}
\begin{picture}(0,0)(0,0)
%\put(-65,-97){\small ${r}$} \put(-65,-200){\small $r$}
%\put(65,-97){\small $r$} \put(65,-200){\small $r$}
%\put(-135,-67){\small {upper bound on $R$}}
%\put(-135,-170){lower bound}%\shortstack{l\\o\\w\\e\\r\\ \\b\\o\\u\\n\\d}}%{%\begin{rotate}{\small ${r}$}\end{rotate}}
%\put(0,-67){\small {upper bound on $R$}} \put(0,-170){\small
%{lower bound on $R$}}
\end{picture}
\centerline{\epsfig{figure=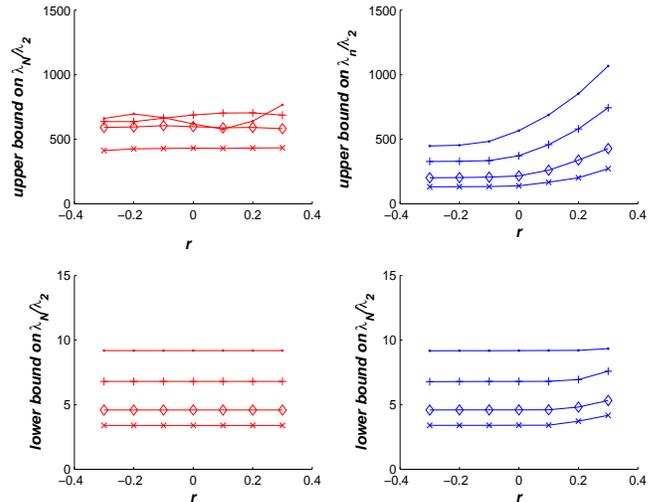,height=6.8cm,width=0.48\textwidth}}
\caption{\label{fig:bounds2} \small %Comparison of... the
Upper and lower bounds given on the eigenratio
$\lambda_N/\lambda_2$ as function of $r$ (as defined in [11]) for
different values of $\gamma$: $\gamma = 2.5 (\cdot)$ , $\gamma = 3
(+)$, $\gamma = 4 (\small{\diamondsuit})$, $\gamma = 5 (\times)$,
$N=10^3$. The lines are guided by the eye. The figure shows the
comparison between the bounds in [2] (on the left) and those
(\ref{OURBOUNDS}) (on the right). The behavior of the actual
eigenratio is plotted in Fig. 1(a).
% the comparison between the bounds in (\ref{MOTTERBOUNDS})
%(on the left) and those (\ref{OURBOUNDS}) (on the right). In the
%inset is shown the behavior of the actual eigenratio.
}
\end{center}
\end{figure}

In \cite{Ni:Mo}
%analytical estimates of theLaplacian eigenvalues are typically used in the literature. For
%example, the
analytical bounds were given to explain the changes observed in
the eigenratio $R$ as the parameter $\gamma$ was varied in a
scale-free network topology. We found that, although these bounds
should hold for any generic network topology, they seem to be
inappropriate to account for the changes in $R$ observed in the
network when degree correlation is introduced. In particular, as
shown in Fig. \ref{fig:bounds2}, the values of the upper bounds on
$R$ computed according to the formulas given in \cite{Ni:Mo} give
estimates which do not reproduce the behavior of the eigenratio
under variation of the network degree correlation.

Therefore we shall seek to define new analytical bounds based on
the mathematical theory of graph spectra. In particular, as
$\lambda_N$ was found to be almost independent from the
correlation coefficient $r$ (see Fig.1(c)), we will focus on
estimating the effects of correlation on the eigenvalue
$\lambda_2$, the parameter known as \emph{algebraic connectivity
of graphs}
\cite{Mohar1}. %In so doing, we will make use of some fundamental
%results  in the field of \emph{algebraic graph theory}.
Specifically, given  a graph, consider a subset of edges which
disconnects the graph in two parts, also termed as a \emph{cut}.
%Isoperimetric problems examine optimal relations between the size
%of the cut and the size of the separated parts.
For a given subset
of vertices, say $S$, we define $h_G(S)$ as the quantity given by:
\begin{equation}
h_G(S)=\frac{{\mathcal D}(S)N}{|S|(N-|S|)},  \label{eq:hG}
\end{equation}
where ${\mathcal D}(S)$ is the number of edges in the boundary of
$S$ and $|S| < \frac{N}{2}$, is the number of vertices in $S$. The
Cheeger
constant of a graph is given by %defined as follows:
$h_G=\min_S h_G(S)$ and the so-called {\em Cheeger inequality}
states that $\lambda_2\leq h_G$ \cite{Cheeg}.
%Remarkably, it can be shown that the following {\em Cheeger
%inequality} holds \cite{Cheeg}. Namely, we have:
%\begin{equation}
%\label{Cheeg1} \lambda_2\leq h_G.
%\end{equation}
%Mohar2

We will show below that \refp{eq:hG} can be successfully used to
compute an upper bound on $\lambda_2$. To overcome the limitations
due to the computation of the subset $S$ that minimizes $h_G(S)$,
we will follow a stochastic approach in order to estimate
$h_G(S)$, starting from the available information we have on the
network. We will assume that the noticeable features of the
network are only the degree distribution and the correlation
specified; all other aspects being completely random. Note that
finding the subset $S$ such as to achieve the minimization of
$h_G(S)$ is known to be an NP-hard problem \cite{Mohar}.

Then, %having fixed the degree sequence,
we can give a full characterization of a randomly chosen subset
$S$ in terms of the number of nodes in it, say $x_i$, belonging to
each class $i$ ($i=1,2...,p$) and the network correlation measure
$r_p$. Analogously let us term as $y_i=\frac{x_i}{n_i}$ the
fraction of nodes in $S$ drawn from each class $i$ ($i=1,2...,p$).
Note that the subset $S$ is not supposed to satisfy any particular
condition, not even of being connected.

Now, we observe that the number of edges in the boundary, say
${\mathcal D}(S)$, is given by the total number of edges starting
from the vertices in $S$, less the ones, say ${\mathcal
I}(S)$, that are contained in $S$, i.e. having both endpoints in $S$. %Let us define the vector
%${\mathbf{x}}={\pmatrix{x_1, x_2, ..., x_p}^{T}}$ representing the
%number of vertices in $|S|$ belonging to each class $i$ and the
%vector ${\mathbf{y}}={\pmatrix{y_1,  y_2, ..., y_p}^{T}}$
%representing the fraction of nodes in $S$ from each of the $p$ classes (i.e. $y_i=\frac{x_i}{n_i}$). %distribution of the nodes in $S$ among thfollowing
Thus we can estimate $\mathcal{I}$ and ${\mathcal D}$ as follows:
\begin{eqnarray*}
& {\mathcal I}(S) = {\mathcal I}({\mathbf{y}}, r_p) =
{\mathcal{E}} {\mathbf{y}}^T E \mathbf{y}, \\
& {\mathcal D}(S)={\mathcal D}({\mathbf{y}}, r_p)={\mathbf{x}}^T
{\mathbf{k}} -2 {\mathcal I}(S)= {2 \mathcal{E} (\mathbf{y}}^T
{\mathbf{q}} - {\mathbf{y}}^T E \mathbf{y}),
\end{eqnarray*}
where ${\mathbf{x}}={\pmatrix{x_1,  x_2, ..., x_p}^{T}}$,
${\mathbf{y}}={\pmatrix{y_1,  y_2, ..., y_p}^{T}}$ and $\mathcal
E$ is the total number of edges in the network.
%represents the fraction of nodes in $S$ from each of the $p$ classes. %distribution of the nodes in $S$ among the $p$ classes; $\rho=\sum_i{k_i}/N$ is the average density of edges within $S$. %$\mathbf{m}=\pmatrix{ m_1 \cr m_2 \cr ...\cr m_p}$.

Thus, $h_G(S)$ becomes:
\begin{equation}
\label{hGS} h_G({\mathbf{y}}, r_p)=\frac{{2 \mathcal{E}
(\mathbf{y}}^T {\mathbf{q}} - {\mathbf{y}}^T E
{\mathbf{y}})N}{({\mathbf{n}}^T {\mathbf{y}}) (N-{\mathbf{n}}^T
{\mathbf{y}})},
\end{equation}
under the constraint that ${\mathbf{n}}^T {\mathbf{y}} <N/2$,
where ${\mathbf{n}}$ is the vector $\pmatrix{n_1,  n_2, ...,
n_p}^{T}$. A numerical optimization algorithm can then be used to
find the subset $S$ that minimizes $h_G(\mathbf{y},r_p)$ in terms
of $y_1, y_2, ..., y_p$ (and subsequently $r_p$) and, in turns, an
upper bound for $\lambda_2$.

Also, from \refp{EE} and \refp{hGS}, we get:
\begin{equation}
\label{proc}
\frac{\partial{h_G(S)}}{\partial{r_p}}\propto\frac{\partial{{\mathcal
D}(S)}}{\partial{r_p}}= %&=& %-2 {\mathcal E} \sigma^2_q
%{\mathbf{y}}^T {\mathbf{M}} {\mathbf{y}} =
-2 {\mathcal E}\sigma^2_q {({\mathbf{y}}^T \mathbf{m})}^2 \leq 0.
\nonumber
\end{equation}
Since, for any vector $\mathbf y$, \refp{proc} is satisfied, then
we have that $\frac{\partial{h_G}}{\partial{r_p}} \leq 0$.
Therefore, we can predict analytically that $h_G$ and hence
$\lambda_2$ will be decreasing as the degree correlation is
increased and, as a consequence, the degree eigenratio $\lambda_N
\over \lambda_2$ will increase for higher values of the
correlation coefficient.

As shown in Fig. \ref{fig:bounds2}, this is indeed what is
observed with the new bound on $\lambda_2$ giving a much better
estimate of the behavior of the eigenratio with respect to both
changes in the degree distribution and the degree correlation.
(Note that $\lambda_N$ is found to be almost independent from $r$
in Fig. \ref{fig:R}.)

\begin{figure}[t]
\begin{center}
\epsfig{width=0.45\textwidth, file=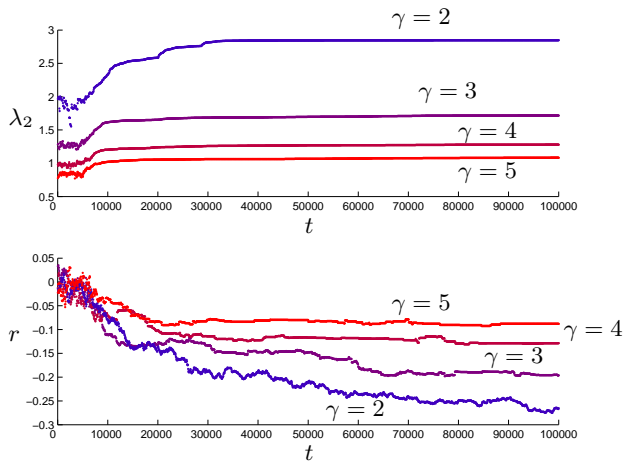}
\begin{picture}(0,0)(0,0)
\put(-220,38){\small ${r}$} \put(-220,120){\small $\lambda_2$}
\put(-108,-7){\small ${t}$} \put(-108,79){\small $t$}
\put(-100,10){\small $\gamma=2$} \put(-40,30){\small $\gamma=3$}
\put(-10,40){\small $\gamma=4$} \put(-75,50){\small $\gamma=5$}
\put(-50,100){\small $\gamma=5$} \put(-50,115){\small $\gamma=4$}%blocco sup
\put(-65,130){\small $\gamma=3$} \put(-75,158){\small $\gamma=2$}
\end{picture}
\end{center}
%\centerline{\bf (a) \hspace{5.5cm} (b)}
\caption{\label{fig:annealing} \small Effects of the maximization
of $\lambda_2$ on the correlation parameter $r$ defined in [11]
for different scale-free networks with $\gamma=[2,3,4,5]$.}
\end{figure}

Another interesting inequality in spectral geometry is due to
Mohar \cite{Mohar}:
\begin{equation}
\label{eqM} \lambda_2 \geq k_{max}-\sqrt{{k_{max}}^2-{h'_{G}}^2},
\end{equation}
where $h'_G=\min_{S}{\frac{{\mathcal D}(S)}{|S|}}={2 \mathcal{E}
(\mathbf{y}}^T {\mathbf{q}} - {\mathbf{y}}^T E
{\mathbf{y}})/({\mathbf{n}}^{T} \mathbf{y})$. Using (\ref{eqM}),
we can also get a lower bound on $\lambda_2$. Then following an
approach similar to the one used to compute the upper bound, it is
easy to show that the lower bound in (\ref{eqM}) has to decrease
with $r$ (note that when making the correlation change, the degree
distribution is fixed and thus, $k_{max}$ cannot vary with $r$).
Then since both the upper and the lower bounds have to decrease
with $r$, $\lambda_2$ is also expected to have the same trend.

In order to derive bounds on $R$, we can use the novel upper and
lower bounds on $\lambda_2$ and the bounds on $\lambda_N$ (that
was observed numerically not to depend on $r$) given in
\cite{Mohar2}:
\begin{equation}
\label{eqBM} \frac{N}{N-1} k_{max}  \leq \lambda_N \leq 2 k_{max}.
\end{equation}
%
%From this relationship and the ones in \refp{Cheeg1} and
%\refp{eqM}
In so doing, we easily get the following analytical bounds on the
Laplacian eigenratio:
\begin{equation}
\label{OURBOUNDS} \label{eqBM} \frac{N}{N-1} \frac{k_{max}}{h_G}
\leq \frac{\lambda_N}{\lambda_2} \leq \frac{2}{1-\sqrt{1-
\frac{h_G'^2}{k_{max}^2}}}.
\end{equation}
The comparison between these bounds computed as explained above
and those proposed in \cite{Ni:Mo} is shown in Fig.
\ref{fig:bounds2}. We observe that the upper bounds in
(\ref{OURBOUNDS}) provide better estimates of changes in the
eigenratio under variations of $\gamma$ and more importantly $r$.

The main result of the derivation presented above is the finding
that disassortative networks synchronize better. Note that such a
derivation can be easily extended to weighted
topologies, i.e. to the case in which %and thus to the case in which the strength of the influence
the strength of the coupling of each vertex on its neighbors is
rescaled by the vertex degree. We wish now to assess whether
negative correlation can be thought of as an emerging property of
networks with an assigned degree distribution in order to improve
their synchronization. Since $\lambda_N$ was found to be
practically independent from variations of $r$,  we use a
simulated annealing meta-heuristic technique \cite{Ki:Ge:Ve}, to
solve the problem of maximising $\lambda_2$ while keeping
unchanged the network degree distribution. Namely, given a network
with a certain degree distribution, we perform the following
iterative procedure. At each iteration $t$, the endpoints of a
randomly selected pair of edges are exchanged if
$\exp{(\frac{-\Delta(\lambda_2)}{T})}>z$; $z$ being an uniformly
distributed random variable between 0 and 1, $\Delta(\lambda_2)$
the variation achieved in the objective function $\lambda_2$
before and after the execution of the move
and $T$ the system \emph{temperature}.%is a control parameter, which by analogy with the original
%formulation of the simulated annealing procedure is known as the
%system 'temperature'. As the algorithm goes on, we make the
%temperature $T$ decrease according to an \emph{exponential cooling
%scheme} (see \cite{Ki:Ge:Ve} for further details).

As shown in Fig. \ref{fig:annealing}, under the effects of this
iterative procedure aimed at maximising $\lambda_2$, we observe
the spontaneous emergence in the network of interest of negative
degree correlation, namely the increasing of its disassortativity.
Thus, following an entirely different approach, we come to the
conclusion that if the degree distribution of a given network is
fixed, then to improve its synchronizability one has to introduce
negative degree correlation among its nodes.% (This might be a
%solution to a classical problem in graph theory optimization which
%is the construction of expander graphs, i.e. highly efficient
%communication networks, characterized by high values of
%$\lambda_2$ and therefore low values of the eigenratio $R$
%\cite{Al:Exp}.)

In conclusion, we have shown that degree correlation among the
nodes of a network of nonlinear oscillators does indeed have an
effect on the Laplacian eigenratio and hence its
synchronizability. Using a novel correlation index, we were able
to show both analytically and numerically that the eigenratio is
lower in disassortative networks. %Note that, as shown incite{ieeepaper}, this result holds also when the coupling strength
%is rescaled by the degree of the nodes.
Hence we conjectured that
in many evolutionary physical and biological networks of nonlinear
oscillators, negative correlation among nodes can be an emerging
property aimed at facilitating the synchronization process. An
iterative nonlinear optimization technique was used to illustrate
this latter point. \narrowtext

\end{multicols}
\end{document}